# Probability distribution of intensity fluctuations of vortex laser beams in the turbulent atmosphere

V. P. AKSENOV,[1] V. V. DUDOROV,[1] V. V. KOLOSOV[1,2*]

[1]*V.E. Zuev Institute of Atmospheric Optics SB RAS, Tomsk, Russia*
[2]*Tomsk Scientific Center SB RAS, Tomsk, Russia*
*Corresponding author: avp@iao.ru*

**Numerical simulation is used to analyze statistical characteristics of vortex beams propagating in the atmosphere. The cumulative distribution function and the probability density function of intensity fluctuations are compared for Gaussian beams and vortex beams. It is shown that for propagation conditions in the turbulent atmosphere corresponding to weak fluctuations (Rytov parameter much smaller than unity), intensity fluctuations at the axis of the Gaussian beam have the lognormal distribution, whereas the probability density distribution of the radiation intensity fluctuations at the axis of the vortex beams is well approximated by the exponential distribution characteristic of conditions of saturated fluctuations (Rytov parameter much larger than unity).**

## 1. INTRODUCTION

To predict the effect of atmospheric turbulence on the functioning of optical systems in the turbulent atmosphere, it is important to know statistical properties of intensity fluctuations of laser beams used in these systems. The cumulative distribution function (CDF) or the probability density function (PDF) characterizes most completely the statistics of these fluctuations (scintillations). It is a good idea to have such a theoretical distribution that simulates intensity fluctuations in the plane of receiving aperture under all possible propagation conditions with the minimal error.

Usually, the Rytov parameter $\beta_0^2 = 1.23 C_n^2 k^{7/6} z^{11/6}$ serves a measure of turbulent distortions of a wave beam. This parameter depends on the structure characteristic of refractive index $C_n^2$ – path length $z$, and wave number of the propagating radiation $k$ [1]. Depending on the value of this parameter, the propagation conditions can be divided into the conditions of weak fluctuations (weak turbulence) when $\beta_0^2 \leq 0.3$, conditions of saturated fluctuations (saturated turbulence) when $\beta_0^2 \geq 10$, and intermediate conditions corresponding to $\beta_0^2$ in the range $\beta_0^2 \approx 5$, which are referred to as the case of strong focusing (moderate turbulence). The combination of conditions of strong focusing and saturated fluctuations is called the case of strong fluctuations.

The probability density function PDF is studied quite intensely as can be seen from the literature [1–8]. The currently proposed theoretical models include the lognormal distribution, exponential one, K distribution, lognormal distribution modulated by the exponential distribution, lognormal distribution modulated by the Rician distribution (also known as the Beckman distribution), and gamma-gamma distribution (gamma distribution modulated by the gamma distribution). Some of these distributions were proposed for different intensity of turbulence: from weak to saturated.

The lognormal and gamma-gamma distributions are now used most widely, according to [4]. These distributions were checked with the results of numerical simulation of propagation of the Gaussian laser beam through statistically homogeneous and isotropic turbulence. The objects under study were the laws of distribution of intensity fluctuations at the beam axis or the laws of distribution of irradiance fluctuations at a receiving aperture of finite size. The proposed distribution models were repeatedly checked experimentally [3, 5, 7–9, 10].

Despite the significant number of theoretical PDF models, they do not work well for a large number of propagation conditions occurring in the atmosphere. Thus, for example, the lognormal distribution significantly understates the frequency of occurrence of giant irradiance spikes observed in the case of strong fluctuations [10]. Some kinds of the distributions include fitting parameters, which are not directly related to values of measured statistical characteristics of irradiance fluctuations, such as the mean intensity and the variance of irradiance fluctuations. These characteristics are quite amendable to calculation based on the existing theories of laser beam propagation in the turbulent medium.

As far as we know, the laws of probability distribution of irradiance fluctuations over the beam cross section remain unstudied.

These studies become urgent in connection with application of beams of special kind and some exotic beams [11]. The emphasis is on vortex beams having the orbital angular momentum [12, 13]. In this paper, the numerical simulation is used to compare the probability distribution functions of intensity fluctuations of the vortex Laguerre–Gaussian beam and fundamental Gaussian beam. The probability distribution functions of two beams at different points of their cross section are examined. The numerical results are compared with the analytical models (lognormal, exponential, gamma-gamma distributions) [4] used most widely by specialists in wave propagation through randomly inhomogeneous media. In addition, the probability distribution densities obtained from the numerical simulation of propagation of optical beams in the atmosphere are approximated by the gamma distribution [14–16] and the fractional exponential distribution proposed in this paper.

A prerequisite for this study was the result [17], which suggests that the scintillation index of the Laguerre–Gaussian beam corresponds to the case of saturated fluctuations at the axis and periphery of the beam and to the case of weak fluctuations at the maximum of the mean intensity.

## 2. NUMERICAL MODEL

The propagation of laser beams was simulated through solution of the parabolic wave equation [18]. Atmospheric turbulence was represented by a set of phase screens [19–21]. The simulation algorithms were organized in the same way as the algorithms of [13, 17]. We used the modified Andrews spectrum of refractive index fluctuations [4], which had the following form

$$\Phi_n(\boldsymbol{\kappa}_\perp, 0) = 0.033 C_n^2 \frac{\exp\left(-\kappa_\perp^2/\kappa_a^2\right)}{\left(\kappa_\perp^2 + \kappa_0^2\right)^{11/6}} \times \left[1 + 1.802 \frac{\kappa}{\kappa_a} - 0.254 \left(\frac{\kappa}{\kappa_a}\right)^{7/6}\right], \quad (1)$$

where $\kappa_0 = 2\pi/M_0$, $\kappa_a = 3.3/m_0$, $m_0$ and $M_0$ are the inner and outer scales of atmospheric turbulence. In the calculations (unless otherwise stipulated), the outer scale of atmospheric turbulence was assumed to be $M_0 = 20a$ and the inner scale was taken $m_0 = 0.08a$ ($a$ is the beam radius). The turbulence was considered isotropic and homogeneous. The turbulent conditions of propagation at the path were specified with the parameter $\beta_0^2$. From the simulated array of random realizations ($N = 5\,000$) of the wave field $E_i(\mathbf{r}, z)$, a sample (empirical) distribution function of random intensity fluctuations was constructed by the standard equation [22]

$$F_N(I) = \frac{1}{N} \sum_{i=1}^{N} \theta(I - I_i), \quad (2)$$

where $\theta(I)$ is the Heaviside function, $I_i(\mathbf{r}, z) = |E_i(\mathbf{r}, z)|^2$ are random values of intensity. $F_N(I)$ is an approximation of the cumulative distribution function $F(I)$. At the known probability density function of intensity fluctuations $P(I)$, the cumulative distribution function can be calculated as [4, 22]

$$F(I) = \int_0^I P(x) dx. \quad (3)$$

The probability density of intensity fluctuations was approximated through the construction of a histogram with the use of a smoothing procedure.

We have studied the statistical characteristics of intensity fluctuations of the Laguerre–Gaussian beam $\mathrm{LG}_0^l$ with the initial field distribution

$$E(r, \theta, z = 0) = \left(\sqrt{2}\frac{r}{a}\right)^{|l|} \exp\left(-\frac{r^2}{a^2}\right) \exp[il\theta], \quad (4)$$

where $r = \sqrt{x^2 + y^2}$ and $\theta = \arctan(y/x)$ are the polar coordinates, $l$ is the topological charge. Equation (4) describes the Gaussian beam (G) if $l = 0$ and the circulation mode $\mathrm{LG}_0^1$ of the Laguerre–Gaussian beam if $l = 1$.

## 3. RESULTS OF NUMERICAL SIMULATION

The calculated probability density function and the cumulative distribution function of irradiance fluctuations obtained in the numerical experiment (solid curves) are shown in Figs. 1–10. For comparison with the results of numerical simulation, we used the known analytical models of the probability density functions, including the lognormal distribution

$$P_{LN}(I) = \frac{1}{I(r)\xi\sqrt{2\pi}} \exp\left[-\left(\ln I(r) - \mu\right)^2 / 2\xi^2\right], \quad (5)$$

where $\langle I(r) \rangle$ – is the mean intensity, angular brackets denote averaging over an ensemble of medium realizations, where

$$B_I(r) = \langle I(r)^2 \rangle - \langle I(r) \rangle^2 \quad (6)$$

is the variance of intensity fluctuations,

$$\sigma_I^2(r) = \frac{B_I(r)}{\langle I(r) \rangle^2} \quad (7)$$

– is the relative variance of intensity fluctuations (scintillation index),

$\mu = \ln\left(\frac{\langle I(r) \rangle}{\sqrt{1 + \sigma_I^2}}\right)$, $\xi^2 = \ln\left(1 + \sigma_I^2\right)$. Lognormal PDF (5) is used most often for the conditions of weak turbulence.

Exponential PDF

$$P_E(I) = \frac{1}{\langle I(r) \rangle} \exp\left(-\frac{I}{\langle I(r) \rangle}\right) \quad (8)$$

is a PDF model for the conditions of saturated fluctuations ($\sigma_I^2(r) \to 1$).

The gamma-gamma PDF is commonly believed a versatile model of distribution suitable for the entire range of turbulent conditions [4, 24] (at least, for a point receiver of radiation [9])

$$P_{GG}(I) = \frac{2(\alpha\beta)^{(\alpha+\beta)/2}}{\Gamma(\alpha)\Gamma(\beta)} I_n^{(\alpha+\beta)/2-1} K_{\alpha-\beta}\left(2\sqrt{\alpha\beta I_n}\right), \quad (9)$$

where $I_n(r) = \frac{I(r)}{\langle I(r) \rangle}$, $\Gamma(x)$ is the gamma function, $K_\nu(x)$ is the second-kind modified Bessel function, $\alpha$ and $\beta$ are the PDF parameters, effective numbers of large-scale and small-scale scatterers, respectively [23]. These parameters are connected to the scintillation index $\sigma_I^2(r)$ through the following equation

$$\sigma_I^2(r) = \frac{1}{\alpha} + \frac{1}{\beta} + \frac{1}{\alpha\beta}. \quad (10)$$

Advantages of PDF model (9) and results of its use for the plane and spherical waves can be found in [4, 24]. It should be noted, however, that the use of analytical estimates of $\alpha$ and $\beta$ [4] not always leads to the sufficiently close agreement between distribution (9) and the distribution obtained from numerical simulation. Therefore, the additional correction of these parameters is necessary. The gamma model of PDF (m-distribution) [14–16]

$$P_G(I) = \frac{m^m I^{m-1}}{\Gamma(m) \langle I \rangle^m} \exp\left(-\frac{mI}{\langle I \rangle}\right) \quad (11)$$

with the two parameters $m$ and $\langle I \rangle$ serves as a basis for construction of the gamma-gamma model. The gamma distribution was initially developed for approximation of the probability density of the amplitude of a wave field scattered by a rough surface [14]. Then it was used for description of statistical properties of a speckle field [15]. In [16], in combination with the exponential distribution, it was also used for approximation of PDF of intensity of a speckle field passed through atmospheric turbulence. As was shown by our calculations (presented below), model (11) can be used for description of statistical properties of the intensity of radiation passed through the turbulent atmosphere and without combination with other distributions. For this purpose, using distribution (11) and equations (6) and (7), we calculate the scintillation index and find the value of the parameter $m$

$$m = \frac{\langle I(r) \rangle^2}{B_I(r)} = \frac{1}{\sigma_I^2(r)}. \quad (12)$$

### 3.1. Weak turbulence

The results of calculation of the probability density function of intensity fluctuations obtained in the numerical experiment (solid curves) for the conditions of weak turbulence at a path with $\beta_0^2 = 0.1$ are depicted in Figs. 1–3. These distributions were obtained at the distance $z = 0.1 z_d$, $z_d = 0.5 k a^2$ – diffraction Rayleigh length. In these figures, the dotted line corresponds to lognormal probability density (5), whereas the dashed curves are drawn according to Eqs. (11)–(12).

Figures 1–3 show the results for the points in the beam cross section, where the scintillation index is smaller than or approximately equal to unity. It follows from Eqs. (11)–(12) that the gamma distribution increases infinitely as the intensity tends to zero, when $\sigma_I^2(r) > 1$ ($m < 1$). That is why for the cases $\sigma_I^2(r) \geq 1$ (Figs. 2a, b) it was taken $m = 1$ in Eq. (11). In this case, the gamma distribution transforms into exponential distribution (8), for which $\sigma_I^2(r) = 1$.

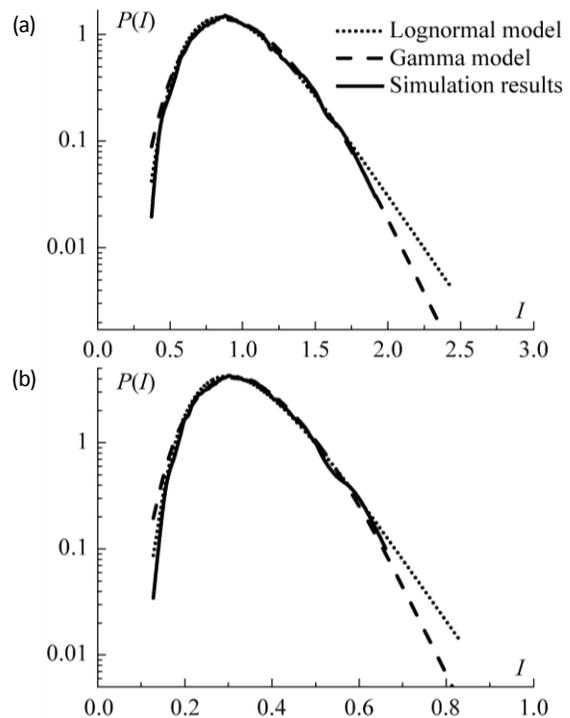

Fig. 1. Probability density function: at the axis ($r = 0$) of the Gaussian beam (equation (4) with $l = 0$) (a), at the point of maximum ($r = 0.47a$) of the mean intensity distribution of the Laguerre–Gaussian beam (equation (3) with $l = 1$) (b). The values of $\sigma_I(r)$ are 0.31 and 0.30, respectively.

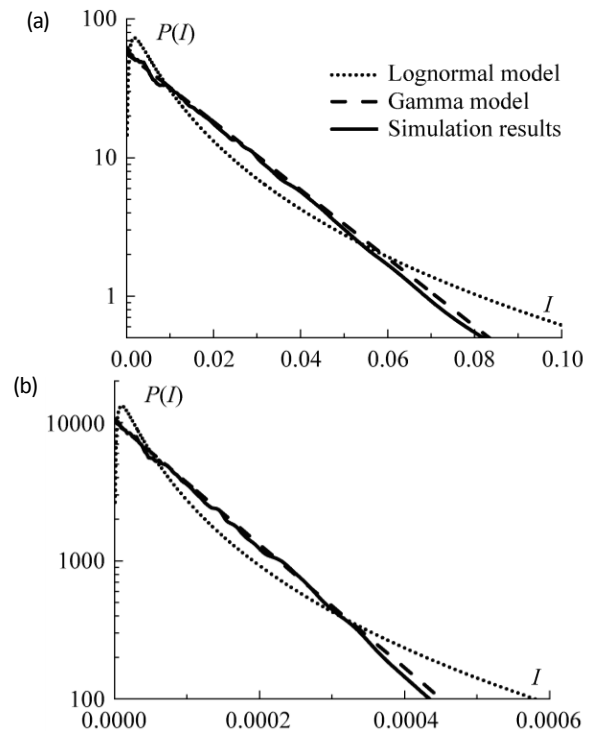

Fig. 2. Probability density function: at the axis ($r = 0$) Laguerre–Gaussian beam (a); at the periphery (at the point $r = 1.9a$) of the Laguerre–Gaussian beam (b). The values of $\sigma_I(r)$ are 1.04, and 1.02, respectively.

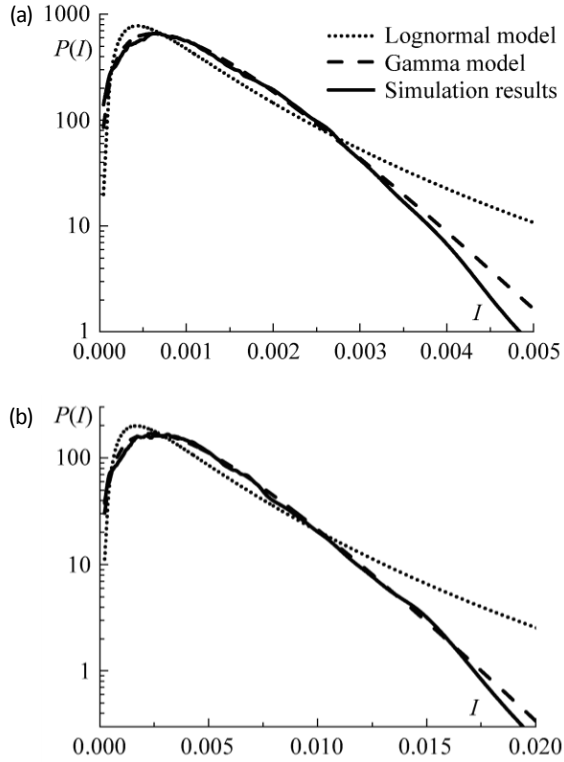

Fig. 3. Probability density function: for the Gaussian beam at the point $r = 1.9a$ (a), for the Laguerre–Gaussian beam at the point $r = 1.4a$ (b). The values of $\sigma_I(r)$ are 0.67 and 0.68, respectively.

It can be seen from the above results that for the conditions of radiation propagation in the turbulent atmosphere corresponding to $z = 0.1 z_d$, and $\beta_0^2 \ll 1$, intensity fluctuations at the axis of the Gaussian beam satisfy the condition $\sigma_I^2(r) \ll 1$ and have the lognormal distribution (Fig. 1a). This is in a good agreement with the theoretical and experimental results [4, 24, 25]. We obtain the analogical result at the ring of the Laguerre–Gaussian beam (Fig. 1b), where its mean intensity is maximal and the condition $\sigma_I^2(r) \ll 1$ is fulfilled for the scintillation index. We can also see that the probability density function for the gamma distribution is very close to the lognormal distribution [16] and can be used for approximation of the calculated results.

At the same time, the scintillation index at the axis of the vortex beam satisfies the condition $\sigma_I^2(r) \approx 1$, and PDF is well approximated by the exponential distribution (Fig. 2a) characteristic of the case of strong fluctuations ($\beta_0^2 \gg 1$). The analogical conditions are also fulfilled at the periphery of the Laguerre–Gaussian beam.

At the points, where the scintillation index of the Laguerre–Gaussian beam and the other considered beams has intermediate values ($0 < \sigma_I^2(r) < 1$), the probability density of intensity fluctuations of these beams is well approximated by the gamma distribution (Figs. 3) with the parameter $m$ inversely proportional to the scintillation index.

Thus, we can draw the conclusion that for the considered beams at the cross-sectional points, for which the condition $0 < \sigma_I^2(r) \leq 1$ is fulfilled, the statistics of intensity fluctuations can be approximated by the gamma distribution.

### 3.2. Moderate turbulence

As was already mentioned, the approximation in the form (11)–(12) becomes inapplicable when the scintillation index $\sigma_I^2(r) > 1$, because in this case the probability density becomes infinite for zero values of $I(r)$. Therefore, it is necessary to use a different approximation, which, as in Eq. (11), transforms to model (8) if $\sigma_I^2(r) = 1$.

It turns out that the PDF model can be represented by the function

$$P_F(I) = \frac{\Gamma(2/m)}{\Gamma^2(1/m)} \frac{m}{\langle I \rangle} \exp\left[-\left(\frac{\Gamma(2/m)}{\Gamma(1/m)}\right)^m \left(\frac{I}{\langle I \rangle}\right)^m\right], \quad (13)$$

with the parameter $m$, which can be found from the following equation:

$$\sigma_I^2(r) + 1 = \frac{\Gamma(1/m)\Gamma(3/m)}{\Gamma^2(2/m)}. \quad (14)$$

If the relative variance of intensity fluctuations takes the value equal to unity ($\sigma_I^2(r) = 1$), then from equation (14) we obtain that $m = 1$ and equation (13) transforms into exponential distribution (8). For $\sigma_I^2(r) > 1$, the parameter $m < 1$. In this case, the higher $\sigma_I^2(r)$, the smaller the parameter $m$. In particular, for the scintillation index $\sigma_I^2(r) = 2$, the parameter $m \approx 0.557$. We call model (13)–(14) the fractional exponential distribution.

The distributions shown in Fig. 4 are obtained at the distance $z = 0.1 z_d$, for the conditions close to moderate turbulence at the path ($\beta_0^2 = 1$). Here, the solid curves are drawn based on the results of numerical simulation, the dashed curves correspond to exponential probability density (8), and the dotted curves are calculated by equations (12)–(13).

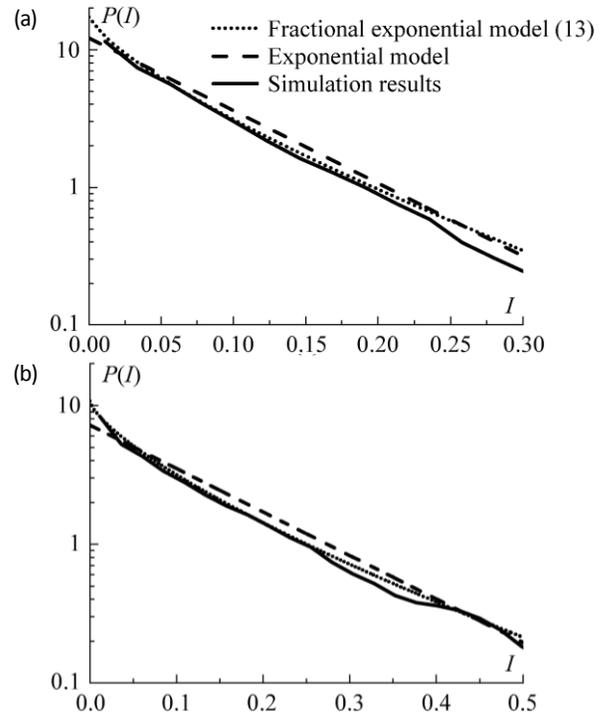

Fig. 4. Probability density function: for the Gaussian beam ($l = 0$) at the point $r = 1.2a$ (a), for the Laguerre–Gaussian beam ($l = 1$) at the point $r = 0.08a$ (b). Scintillation index $\sigma_I(r) = 1.17$ (a) and $\sigma_I(r) = 1.19$ (b), parameter $m = 0.748$ (a) and $m = 0.724$ (b).

It can be seen that at $\sigma_I(r) > 1.1$ the probability density distributions obtained in numerical calculations (solid curve) become different from the exponential distribution (dashed curve) and can be well approximated by the fractional exponential distribution (dotted curve).

Figure 5 depicts the results of calculation of intensity fluctuations of the Laguerre–Gaussian beam for the distance $z = 0.1 z_d$ and conditions of moderate turbulence at the path ($\beta_0^2 = 1.0$). With an increase of $\sigma_I(r)$, the difference of the distribution density obtained in the numerical calculations from the exponential distribution increases.

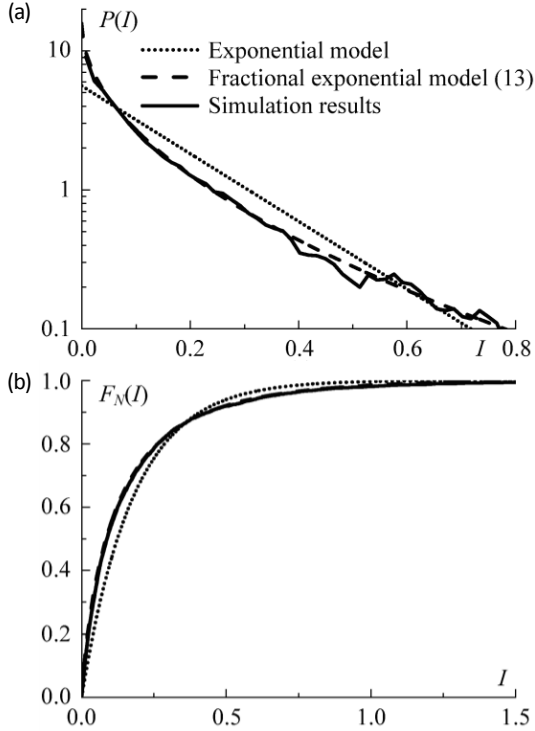

Fig. 5. Probability density function (a) and the cumulative distribution function (b) of intensity fluctuations of the Laguerre–Gaussian beam at the point $r = 1.08a$: $\sigma_I(r) = 1.50$, $m = 0.514$, $z = 0.1 z_d$. Numerical simulation (solid curves), exponential probability density (8) (dotted curves), and fractional distribution (13)–(14) (dashed curves).

Line designations in Fig. 5a are analogous to Fig. 4. In Fig. 5b, the solid curve is the sample distribution (2). The dashed curve is cumulative distribution function (CDF) (3) corresponding to probability density (13)–(14). The dotted line corresponds to the exponential probability density.

One can see that approximation (13)–(14) provides a satisfactory agreement for the probability density function and the cumulative distribution function.

Figures 6–7 show the results of comparison of the numerical calculations with the lognormal distribution, gamma-gamma distribution, and fractional exponential distribution. The calculations were performed for the distribution function and probability density of intensity fluctuations of the Laguerre–Gaussian beam at different distances from the beam center for conditions of moderate turbulence and the atmospheric path length equal to the Rayleigh diffraction length. From here on, the parameters of gamma-gamma distribution (9) $\alpha$ and $\beta$ were found from fitting to the numerically obtained distribution. It can be seen that CDF of intensity fluctuations is well approximated by the gamma-gamma distribution and the fractional exponential distribution, which give close results in this case. However, the probability density functions for these distributions behave differently, especially, in the zone of low intensity values. Fractional distribution (13)–(14) tends to a finite value, while log-normal (5) and gamma-gamma (9) distributions tend to zero as $I \to 0$. For the log-normal distribution, the difference from the numerical results is large as compared to the other distributions.

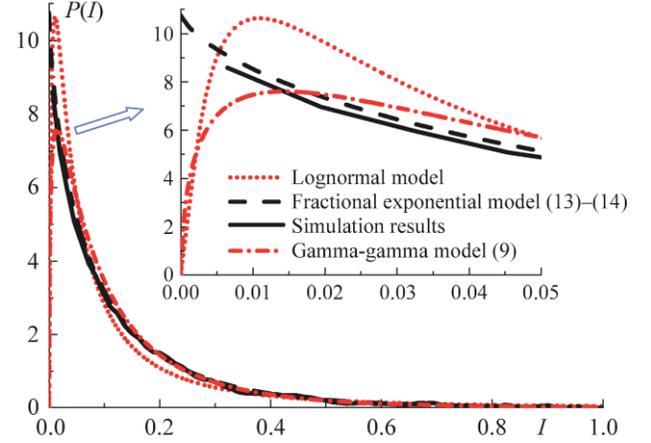

Fig. 6. Probability density function of intensity fluctuations of the Laguerre–Gaussian beam at the point $r = 0.08a$: $\sigma_I(r) = 1.19$, $z = z_d$. Solid curve – results of numerical simulation, dash and dot curve – gamma-gamma distribution (9) ($\alpha = 1.7950$, $\beta = 1.7935$), dots – lognormal distribution, dashed curve – approximation (13)–(14) ($m = 0.724$). Enlarged parts of the corresponding curves are shown in insets.

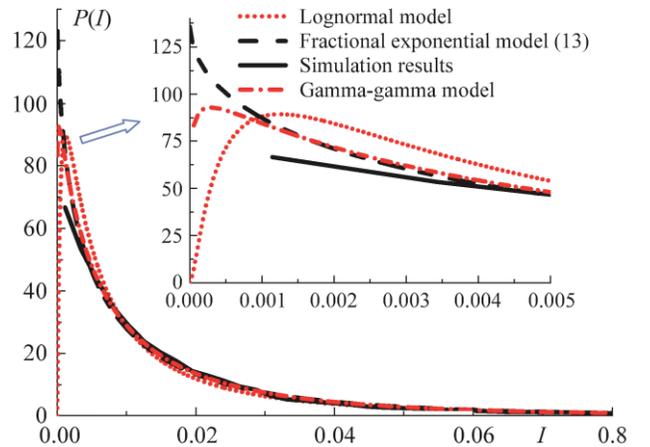

Fig. 7. Probability density function of intensity fluctuations of the Laguerre–Gaussian beam at the point $r = 0.133a$: $\sigma_I(r) = 1.42$, $z = z_d$. Solid curve – results of numerical simulation, dash and dot curve – gamma-gamma distribution (9) ($\alpha = 1.3620$, $\beta = 1.3617$), dotted curve – lognormal distribution, dashed curve – approximation (13)–(14) ($m = 0.555$). Enlarged parts of the corresponding curves are shown in insets.

## 3.3. Strong turbulence

Figure 8 shows the results of calculation of intensity fluctuations of the Gaussian beam for the distance $z = 0.1z_d$ and conditions of strong turbulence at the path ($\beta_0^2 = 10.0$). The results are given for the points in the beam cross section, for which the scintillation index exceeds the value $\sigma_I^2(r) > 3.0$.

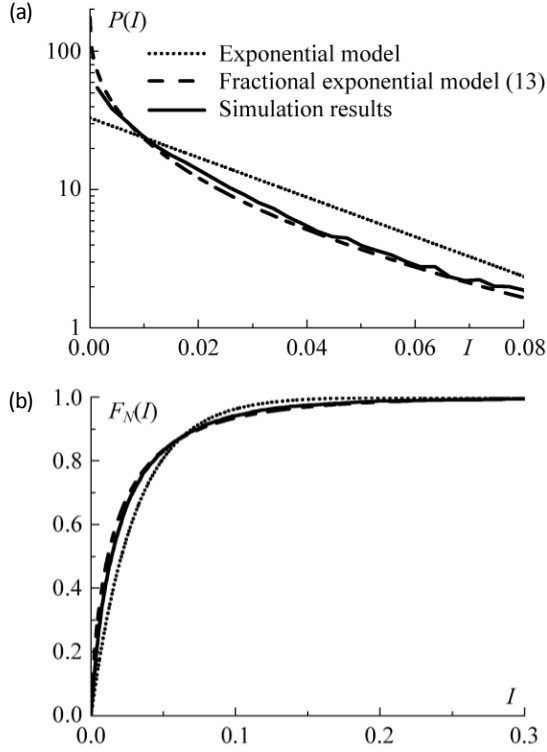

Fig. 8. Probability density function (a) and cumulative distribution function (b) of intensity fluctuations of the Gaussian beam at the point $r = 1.9a$: $\sigma_I(r) = 1.81$, $m = 0.403$, $z = 0.1z_d$. Solid curves – numerical simulation, dotted curves – exponential probability density (8), and dashed curves – fractional distribution (13)–(14).

Figures 8–10 depict the results of calculation of the distribution function and the probability density of intensity fluctuations of the Gaussian beam at different distances from the beam center for the conditions of strong turbulence and different diffraction conditions. In Fig. 8, the numerical calculations are compared with the exponential and fractional exponential distributions. In Figs. 9–10, the numerical results are compared with the fractional exponential, lognormal, and gamma-gamma distributions.

It can seen that under conditions of strong turbulence the distribution laws of intensity fluctuations do not vary qualitatively over the beam cross section. The parameters $\alpha$ and $\beta$ of gamma-gamma distribution $P_{GG}(I)$ (9) are found here through the fitting to the numerically determined distribution. However, it can be shown that the applicability condition of distribution (9) provides that these parameters should satisfy the conditions $\alpha \geq 1$, $\beta \geq 1$. If these conditions are not fulfilled, then we come to the result contradicting, in our opinion, the physics of the considered phenomena: $P_{GG}(I) \to \infty$ if $I \to 0$. This local singularity of $P_{GG}(I)$ is connected with the similar singularity of gamma distribution (11). In view of these singularities of $P_{GG}(I)$ and calculated values of $\alpha$ and $\beta$, the behavior of the corresponding curves in Figs. 9 and 10 cannot be considered as reliable (especially, for small $I$).

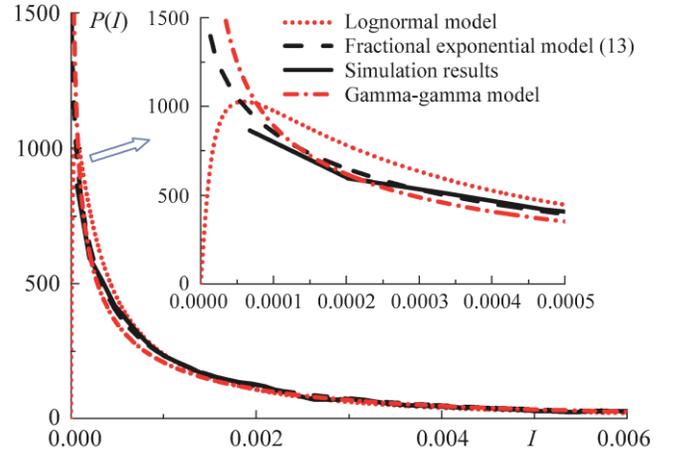

Fig. 9. Probability density function of intensity fluctuations of the Gaussian beam at the point $r = 6.59a$: $\sigma_I(r) = 1.84$, $\beta_0^2 = 10.0$, $z = 10z_d$. Solid curve – results of numerical simulation, dash and dot curve – gamma-gamma distribution (9) ($\alpha = 1.5895$, $\beta = 0.5889$), dotted curve – lognormal distribution, dashed curve – approximation (13)–(14) ($m = 0.395$). Enlarged parts of the corresponding curves are shown in insets.

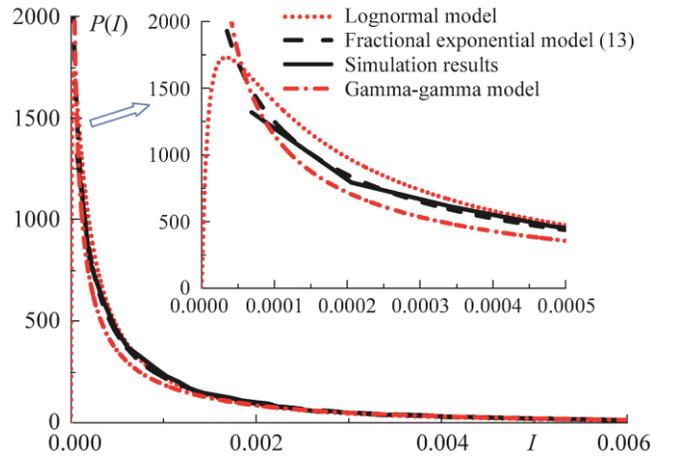

Fig. 10. Probability density function of intensity fluctuations at the periphery of the Gaussian beam at the point $r = 9.52a$: $\sigma_I(r) = 2.11$, $\beta_0^2 = 10.0$, $z = 10z_d$. Solid curve – results of numerical simulation, dash and dot curve – gamma-gamma distribution (9) ($\alpha = 1.4500$, $\beta = 0.4499$), dotted curve – lognormal distribution, dashed curve – approximation (13)–(14) ($m = 0.339$). Enlarged parts of the corresponding curves are shown in insets.

## 4. CONCLUSIONS

The results reported in this paper cover a wide range of turbulent conditions of propagation of various laser beams. The calculations have been performed for the Rytov parameter $\beta_0^2 = 0.1$, 1.0, and 10.0, which correspond to the conditions of weak, moderate, and strong turbulence. The results presented have been obtained for different separations of the observation point from the beam axis (from axial points $r=0$ to points at the beam periphery $r=1.9a$) and different separations of the observation plane from the plane of the emitting aperture ($z=0.1z_d$, $z=1.0z_d$, and $z=10z_d$).

In all the cases, regardless of the beam type, turbulent conditions, and position of the observation point, the statistics of fluctuations (distribution function and probability density of intensity fluctuations) is uniquely determined by the values of the mean intensity and the variance of fluctuations. For fluctuations with the scintillation index $\sigma_I^2(r)<1$, the probability density is satisfactorily approximated by gamma distribution (11), while for fluctuations with the scintillation index $\sigma_I^2(r)>1$ the probability density is satisfactorily approximated by the fractional exponential distribution of form (13).

It should be noted that as the scintillation index tends to unity, both distributions (11) and (13) transform into the exponential distribution. As the scintillation index tends to zero, gamma distribution (11) approaches the lognormal distribution. Both these results correspond to the theoretical results obtained for these limiting cases [4, 7, 24–25].

## REFERENCES

1. S. M. Flatté, C. Bracher, and G. Wang, "Probability density functions of irradiance for waves in atmospheric turbulence calculated by numerical simulation," Opt. Soc. Am. A **11**, 2080–2092 (1994).
2. R. J. Hill and R. G. Frehlich, "Probability distribution of irradiance for the onset of strong scintillation," J. Opt. Soc. Am. A **14**, 1530–1540 (1997).
3. J. H. Churnside and R. G. Frehlich, "Experimental evaluation of log-normally modulated Rician and IK models of optical scintillation in the atmosphere," J. Opt. Soc. Am. A **6**, 1760–1766 (1989).
4. L. C. Andrews and R. L. Phillips, "Laser Beam Propagation through Random Media," Proc. SPIE (2005).
5. J. H. Churnside and R. J. Hill, "Probability density of irradiance scintillations for strong path-integrated refractive turbulence," J. Opt. Soc. Am. A **4**, 727–733 (1987).
6. S. D. Lyke, D. G. Voelz, and M. C. Roggemann, "Probability density of aperture-averaged irradiance fluctuations for long range free space optical communication links," Appl. Opt. **48**, 6511–6527 (2009).
7. J. R. W. Mclaren, J. C. Thomas, J. L. Mackintosh, K. A. Mudge, K. J. Grant, B. A. Clare, and W. G. Cowley, "Comparison of probability density functions for analyzing irradiance statistics due to atmospheric turbulence," Appl. Opt. **51**, 5996–6002 (2012).
8. R. Barrios and F. Dios, "Exponentiated Weibull distribution family under aperture averaging for Gaussian beam waves," Opt. Express **20**, 13055–13064 (2012).
9. F. S. Vetelino, C. Young, and L. Andrews, "Fade statistics and aperture averaging for Gaussian beam waves in moderate-to-strong turbulence," Appl. Opt. **46**, 3780–3790 (2007).
10. S. L. Lachinova and M. A. Vorontsov, "Giant irradiance spikes in laser beam propagation in volume turbulence: Analysis and impact," J. Opt. **18**, 025608 (2016), 12 pp.
11. D. L. Andrews, in Structured Light and its Applications: An Introduction to Phase-Structured Beams and Nanoscale Optical Forces (Academic, 2008).
12. A. M. Yao and M. J. Padgett, "Orbital angular momentum: Origins, behavior and applications," Adv. Opt. Photon. **3**, 161–204 (2011).
13. V. P. Aksenov, V.V. Dudorov, V. V. Kolosov, "Properties of vortex beams formed by an array of fiber lasers and their propagation in a turbulent atmosphere," Kvantovaya Elektronika **46**, 726–732 (2016).
14. M. Nakagami, "The *m* distribution – a general formula of intensity distribution of rapid fading," in Statistical Methods in Radio Wave Propagation, W. C. Hoffman, ed. (Pergamon, New York, 1960), pp. 3–36.
15. J. W. Goodman, "Statistical properties of laser speckle patterns," in Laser Speckle and Related Phenomena, J. C. Dainty, ed. (Springer-Verlag, New York, 1975). Chap. 2.
16. V. S. R. Gudimetla and J. F. Holmes, "Probability density function of the intensity for a laser-generated speckle field after propagation through the turbulent atmosphere," J. Opt. Soc. Am. **72**, 1213–1218 (1982).
17. V. P. Aksenov., V. V. Kolosov, "Scintillations of optical vortex in randomly inhomogeneous medium," Photon. Res. **3**, No. 2, 44–47 (2015).
18. S. M. Rytov, Yu. A. Kravtsov, and V. I. Tatarskii, " Principles of Statistical Radiophysics," Vol. 4. Wave Propagation through Random Media (Springer, 1988).
19. J. A. Fleck, J. R. Morris, and M. D. Feit, "Time-dependent propagation of high-energy laser beams through the atmosphere," Appl. Phys. **10**, 129–160 (1976).
20. P. A. Konyaev and V. P. Lukin, "Thermal distortions of focused laser beams in the atmosphere," Appl. Opt. **24**, 415–421 (1985).
21. J. M. Martin and S. M. Flatté, "Intensity images and statistics from numerical simulation of wave propagation in 3-D random media," Appl. Opt. **27**, 2111–2126 (1988).
22. A.W. van der Vaart (1998), Asymptotic statistics. Cambridge University Press. p. 265; https://en.wikipedia.org/wiki/Empirical_distribution_function
23. M. A. Al-Habash, L. C. Andrews, and R. L. Phillips, "Mathematical model for the irradiance probability density function of a laser beam propagating through turbulent media," Opt. Eng. **40**, 1554–1562 (2001).
24. A. M. Prokhorov, F. V. Bunkin, K. S. Gochelashvily, and V. I. Shishov, "Laser irradiance in turbulent media," Proc. IEEE **63**, 790–809 (1975).
25. M. E. Gracheva, A. S. Gurvich, S. S. Kashkarov, V. V. Pokasov, "Similarity Relations and their Experimental Verification for Strong Intensity Fluctuations of Laser Radiation," in Laser beam propagation in the atmosphere. J. W. Strohbehn, ed. (Springer-Verlag, 1978), pp. 107–128.